\documentclass[prl,floatfix,twocolumn]{revtex4}

\usepackage{times,amssymb,amsmath}
\usepackage{graphicx}
\usepackage{hyperref}

\newcommand{\bfp}{\ensuremath{\boldsymbol{p}}}
\newcommand{\bfq}{\ensuremath{\boldsymbol{q}}}
\newcommand{\bfr}{\ensuremath{\boldsymbol{r}}}

\begin{document} 

\title{A Corpuscular Picture of Electrons in Chemical Bond}

\author{Koji Ando}

\affiliation{Department of Chemistry, Graduate School of Science,
Kyoto University, Sakyo-ku, Kyoto 606-8502, Japan}

\date{\today}

\begin{abstract}
We introduce a theory of chemical bond with a corpuscular picture of electrons. It employs a minimal set of localized electron wave packets with \lq floating and breathing\rq\ degrees of freedom and the spin-coupling of non-orthogonal valence-bond theory. It accurately describes chemical bonds in ground and excited states of spin singlet and triplet, in a distinct manner from conventional theories, indicating potential for establishing a dynamical theory of electrons in chemical bonds.

\end{abstract}

\maketitle

Modern concept of chemical bond has been prevailed by the molecular
orbital (MO) model \cite{SzaboOstlund,McWeeny92} 
that is based on one-electron orbitals with the mean-field approximation. 
Even the recently blooming density functional theory \cite{Parr89}, 
whose genuine
form should be independent of the MO concept, is practically
based on the MO model in the Kohn-Sham scheme.
The MO theory first solves a one-electron 
wave equation in the mean-field approximation, 
and assigns electrons to the resultant MOs. The MOs
reflect the molecular symmetry and are thus delocalized
over the molecule or molecular moieties. They accommodate nodal
structures of orbital phases 
that play dominant roles in determining chemical reactivity,
as unveiled by the Frontier Orbital Theory \cite{Fukui71} 
and the Woodward-Hoffmann Rule \cite{Woodward69}.
Adequacy of the MO model has been endorsed by 
photoelectron \cite{Itatani04,Smirnova09}, electron
scattering \cite{Yamazaki15}, and Penning ionization \cite{Ohno04} spectroscopies, 
some of which even
attempt to observe the MOs (more precisely the Dyson orbitals).

While the MO picture emphasizes the delocalized 
wave picture of one-electron orbitals, 
valence-bond (VB) model offers
an alternative concept \cite{Shaik08,Pauling60}.
The VB theory describes the electronic wave function
as an antisymmetrized product of spatial and spin functions,
the former consisting of
a product of atomic orbitals (AOs). 
The \lq resonance structures\rq\ such as covalent and ionic ones
provide intuitive understanding of chemical bond formation, 
reactivity, environmental effects, and so forth.
In the full configuration-interaction (CI) limit with the 
same AO basis,
the fully spin-coupled VB and
the MO-CI are equivalent.
In many cases, the VB model provides a compact and intuitive
description of chemical phenomena.

Both MO and VB calculations have been based on
linear-combination of atomic orbitals (LCAO)
in which the AOs
are clamped at the nuclear centers. 
By contrast,
an alternative picture we present here is based on a 
\lq floating and breathing\rq\ wave packet (WP) model of corpuscular electrons \cite{Ando09,Ando12}. 
The model not only offers a distinct picture of static chemical bond,
but also has direct connection to the electron WP dynamics,
an emerging arena with the recent advent of atto-second spectroscopic techniques \cite{Itatani04,Smirnova09,Kling08,Krausz09,Kato12,Vacher14,Oliveira15}. 

There exist many previous works 
on the floating orbital model within the MO framework \cite{Frost67,Nakatsuji78}.
Probably 
due to the historical prevalence of MO over VB,
the orbital floating in the VB model has not been much explored, 
with a few exception to our knowledge on H$_2$ molecule
that however concluded negatively on the importance of orbital floating \cite{Shull58,Reeves63}.
Here we demonstrate the contrary for molecules
with more electrons and atoms and for excited states.

We employ an ordinary form of
antisymmetrized product of spatial ($\Phi$) and spin ($\Theta$) functions
for $N$-electron wave functions
\begin{equation}
\Psi(1,\cdots,N)={\cal A}[\Phi(\bfq_{1},\cdots,\bfq_{N})\Theta(1,\cdots,N)],
\end{equation}
in which ${\cal A}$ is the antisymmetriser and $\bfq_i$ represents the
spatial coordinates of electrons.
As usual, the spatial part assumes a product form of one-electron
orbitals,
\begin{equation}
\Phi(\bfq_{1},\cdots,\bfq_{N})=\phi_{1}(\bfq_{1})\cdots\phi_{N}(\bfq_{N}) .
\end{equation}
In the conventional VB method,
the orbitals $\phi_{i}(\bfq_{i})$ 
are constructed
from the LCAO.
By contrast, we employ 
for $\phi_{i}(\bfq_{i})$ 
\lq floating and breathing\rq\ minimal
WPs of a form
\begin{equation}
\phi(\bfq,t)
=(2\pi\rho_{t}^{2})^{-\frac{3}{4}}
\exp[-\gamma_{t}|\bfq-\bfr_{t}|^{2}
+ {i}\bfp_{t}\cdot(\bfq-\bfr_{t})/\hbar] ,
\label{eq:wpbasis}
\end{equation}
in which $\bfr_{t}$ and $\bfp_{t}$ represent the WP center and its momentum,
and $\gamma_{t}=1/4\rho_{t}^{2}-(i/2\hbar)\pi_{t}/\rho_{t}$ where
$\rho_{t}$ and $\pi_{t}$ represent the WP width and its momentum \cite{Arickx86}.
These dynamical variables are
determined from the time-dependent or
independent variational principle \cite{Kuratsuji81,Ando14}. 

The spin part
$\Theta(1,\cdots,N)$
consists of the spin eigenfunctions \cite{McWeeny92}. 
We assume here a simple \lq perfect-pairing (PP)\rq\ form,
\begin{equation}
\Theta =
\theta(1,2)
\cdots\theta(N_{p}-1,N_{p})\alpha(N_{p}+1)\cdots\alpha(N) ,
\end{equation}
in which $\alpha$ and $\beta$ are the one-electron spin functions and
\(
\theta(i,j)=(\alpha(i)\beta(j)-\beta(i)\alpha(j))/\sqrt{2}
\)
is the singlet-pair function. The PP model has
been chosen for simplicity and adequacy for the molecules
studied here.
We denote this model \lq WP-VBPP\rq\ or simply \lq WP-VB\rq.

\begin{figure}[t]
    \centering
        \includegraphics[width=0.45\textwidth]{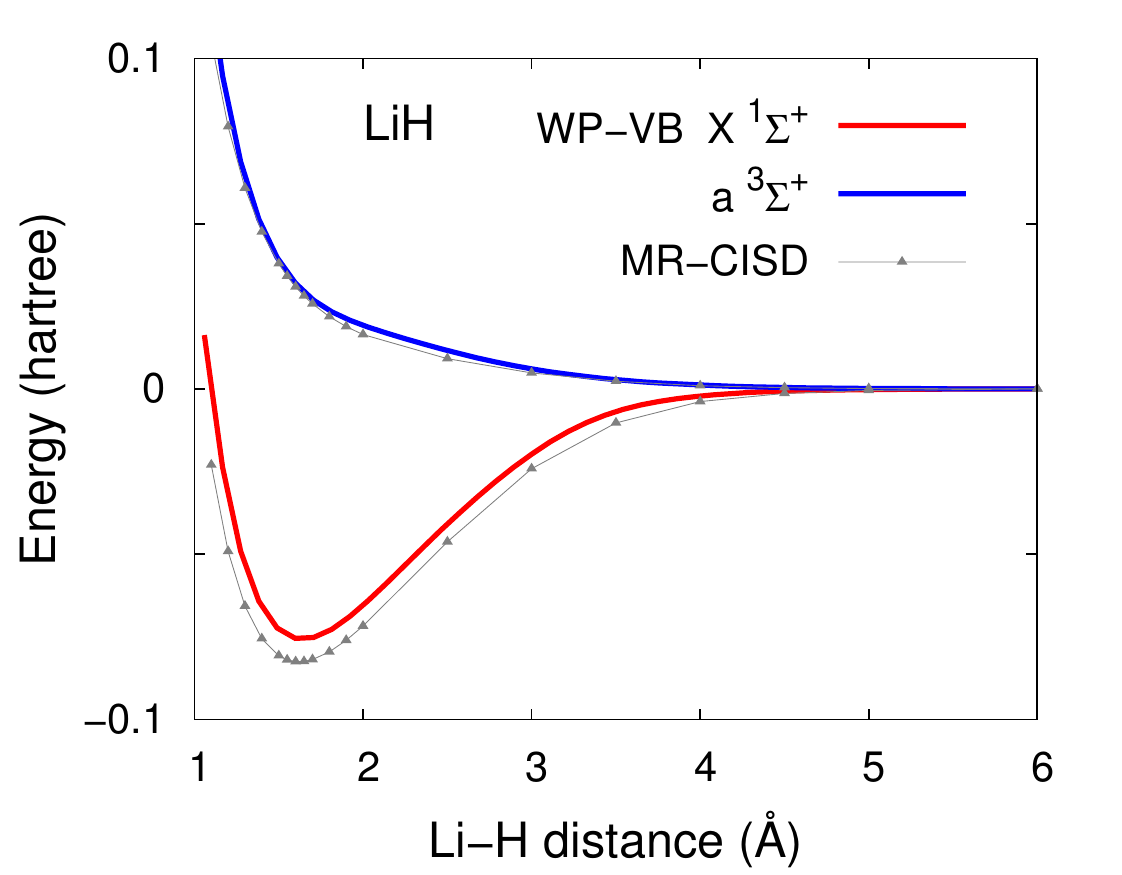}
    \caption{
Potential energy curves of singlet and triplet ground states, $X^{1}\Sigma^{+}$
and $a^{3}\Sigma^{+}$, of LiH.
WP-VB denotes the present floating and breathing minimal 
electron wave packet model with valence-bond spin-coupling.
MR-CISD denotes the state-of-the-art reference calculations from
multi-reference configuration-interaction with single and double excitations,
with the cc-pVDZ basis set consisting of 27 atomic orbitals.
The absolute energies from WP-VB and MR-CISD calculations were shifted 
by +7.109 hartree and +7.932 hartree, respectively, 
to match the dissociation limits. 
}
\end{figure}

To calculate the potential energy surfaces in the ground state
of a given total spin, the energy expectation
$E = \langle \Psi | H | \Psi \rangle / \langle \Psi | \Psi \rangle$
is minimized 
with respect to the center and width variables $\bfr$ and $\rho$
of Eq. (\ref{eq:wpbasis}),
with the momentum variables $\bfp$ and $\pi$ nullified \cite{Tsue92}.
Figure 1 presents the resultant potential energy curves for the singlet $X^{1}\Sigma^{+}$
and the triplet $a^{3}\Sigma^{+}$ states of LiH.
They are compared to the reference calculations with the state-of-the-art wave function approach of
the multi-reference CI with single and double excitations (MR-CISD)
with the cc-pVDZ basis set that consisted of 27 primitive AOs in total.
(We used the program GAMESS \cite{GAMESS} 
for the standard MO-based calculations.)
By contrast,
the WP-VB calculation employed the \lq minimal\rq\ basis, that is, only one 
WP per electron.
The accurate energy curves were obtained by
optimizing both the floating ($\bfr$)
and breathing ($\rho$) degrees of freedom,
which 
accounted for the effects of polarization
and so-called \lq dynamic\rq\ correlation, respectively,
while the VB spin-coupling accounted
for the \lq static\rq\ correlation.

The minimal WP model offers a simple corpuscular picture of electrons
in chemical bonds.
Figures 2(a) and (a) display the electron WPs in the
$X^{1}\Sigma^{+}$ and $a^{3}\Sigma^{+}$ states by
circles with radius of the WP width $\rho$.
They correspond to the electrons of, 
in the ascending order of the radius size,
two Li 1s, H 1s, and Li 2s.
While the former three WPs center around the nuclear positions,
the \lq Li 2s\rq\ WP exhibits notable displacements.
The pictures of the WPs correspond well with
the conventional MOs
in Fig. 2(c) and (d),
two singly-occupied alpha-spin MOs of
the $a^{3}\Sigma^{+}$ state,
from the restricted open-shell Hartree-Fock
calculation with the cc-pVDZ basis set.
The HOMO and LUMO
of the $X^{1}\Sigma^{+}$ state
look similar to them. 
These MOs are described by the linear combination of 27 AOs
with various size and higher angular momenta for polarization.
The corresponding electron distributions are properly described by
the floating and breathing minimal WPs in Fig. 2(a) and (b).

\begin{figure}[t]
    \centering
        \includegraphics[width=0.45\textwidth]{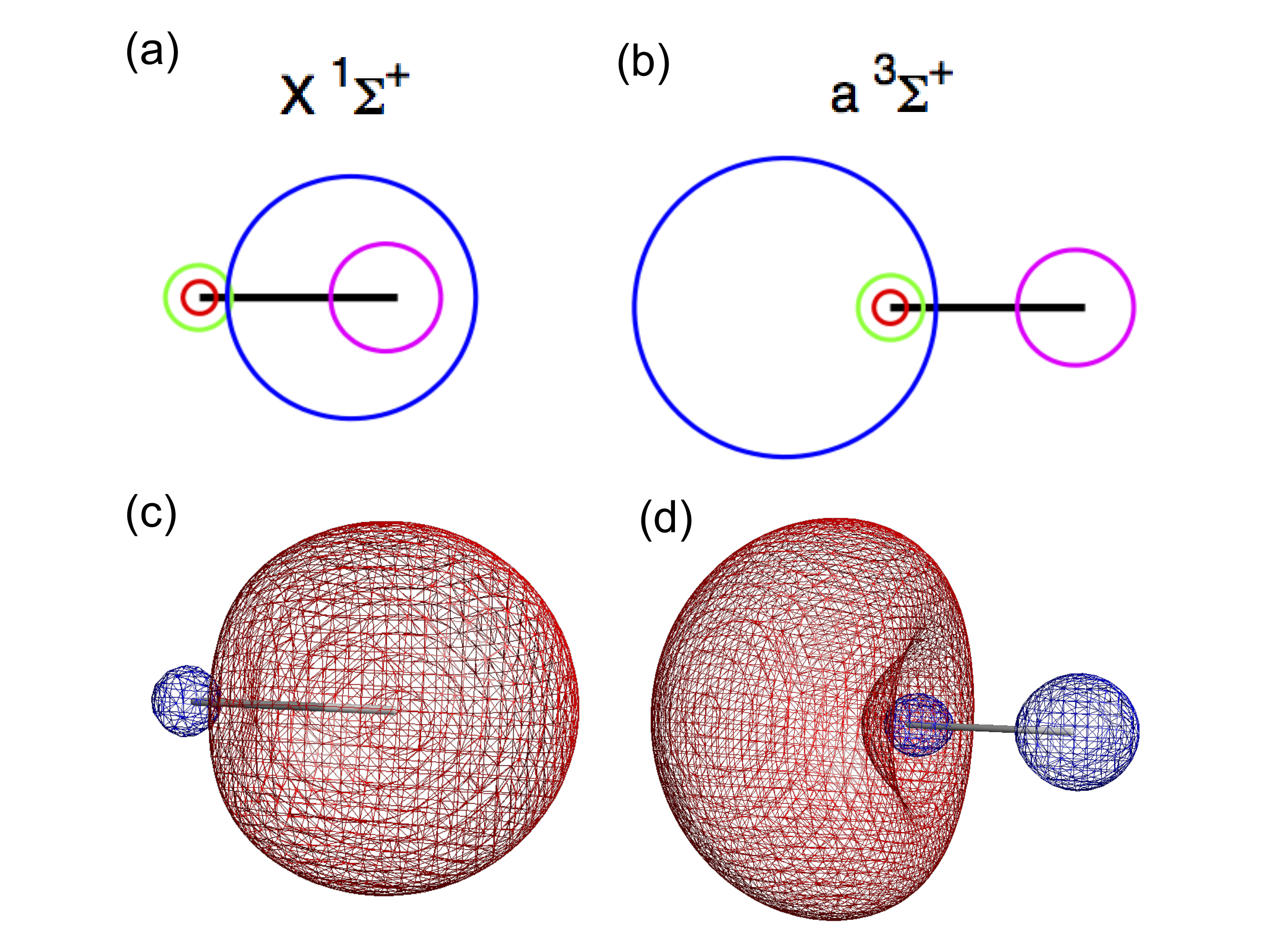}
    \caption{
Electron wave packets compared with molecular orbitals.
(a) $X^{1}\Sigma^{+}$ state. (b) $a^{3}\Sigma^{+}$ state.
In (a) and (b), the radii of circles represent the width ($\rho$) of
electron wave packets that correspond to
Li 1s (red and green), H 1s (magenta), and
Li 2s (blue) electrons.
(C) and (D), two singly-occupied alpha-spin molecular orbitals of $a^{3}\Sigma^{+}$state
of LiH from restricted open-shell Hartree-Fock calculations with
the cc-pVDZ basis set.
}
\end{figure}

Next we examine electronic excited states.
In the MO framework, 
the most standard is the CI method.
Another common approach is the equation-of-motion
or Green function method that gives excitation energies from poles
of electron propagator \cite{McWeeny92}. 
In the conventional VB framework, excited states are computed
with non-orthogonal CI among independent spin-coupling configurations \cite{Shaik08}.
Although this VB-CI method would be applicable with the present
WP-VB scheme, 
we take an alternative route based on the
idea of the propagator theory,
since we have the electron WP dynamics in perspective.

A complete method to construct the electron propagator in the present WP framework
is via the coherent-state path-integral formulation \cite{Kuratsuji81,Klauder84}. 
It is implemented in the initial-value-representation
with the steepest-descent evaluation and Monte Carlo integration 
of the path-integral \cite{Ando14,Kay05}. 
However, here we take a simplified route: 
we construct potential energy surfaces for
the electron WP motion, and solve numerically the time-independent Schr{\"o}dinger
equation 
to obtain electronic excitation energies.
This reduces to one or two dimensional
calculations for the case of LiH as follows. 
From a preliminary
normal-mode analysis, we found that the motion of a particular
electron WP, that corresponds to the Li 2s electron, 
dominates the lowest energy excitations. We thus shifted the center
of that WP along the molecular axis to construct a potential
for that electron motion. Numerical solution of the Schr{\"o}dinger
equation in this potential gives the excitation energies of the $\Sigma$
states. 
Similarly, 
the WP center was shifted perpendicular
to the molecular axis for the $\Pi$ states.

\begin{figure}[t]
    \centering
        \includegraphics[width=0.43\textwidth]{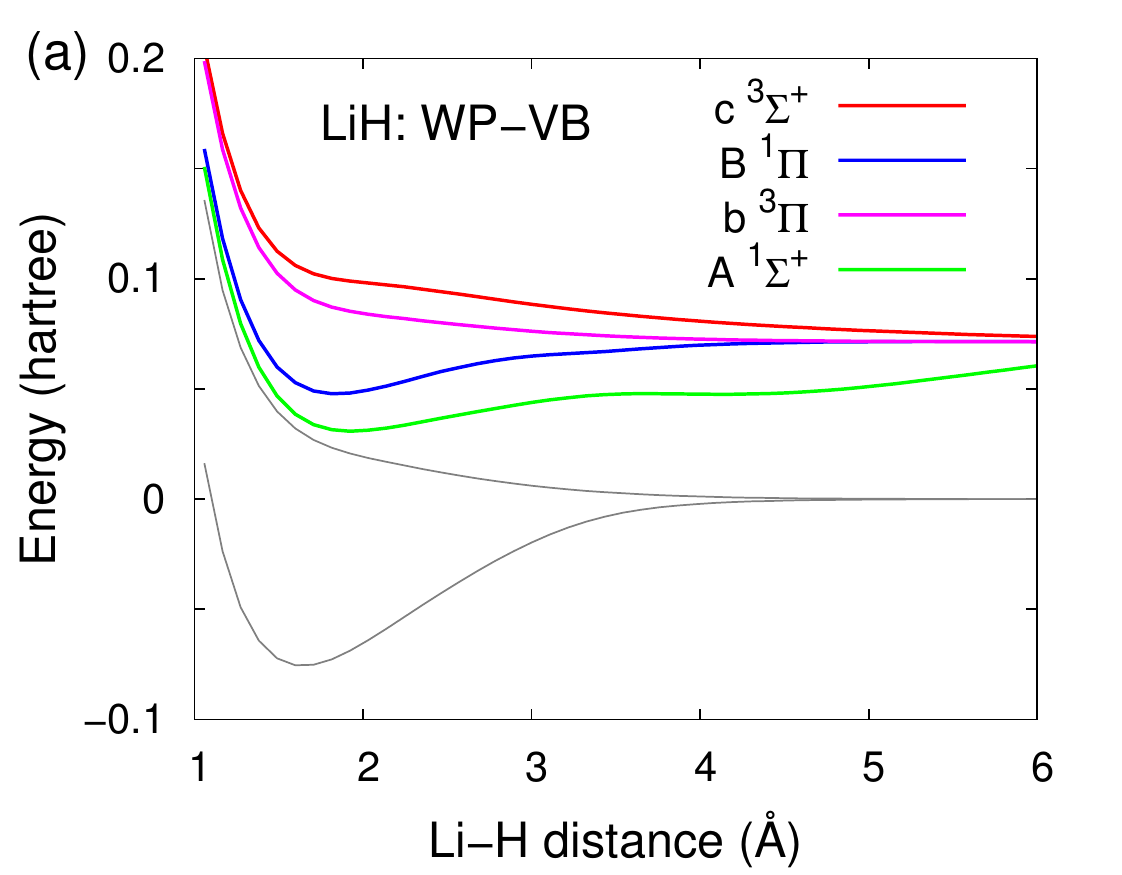}
        \includegraphics[width=0.43\textwidth]{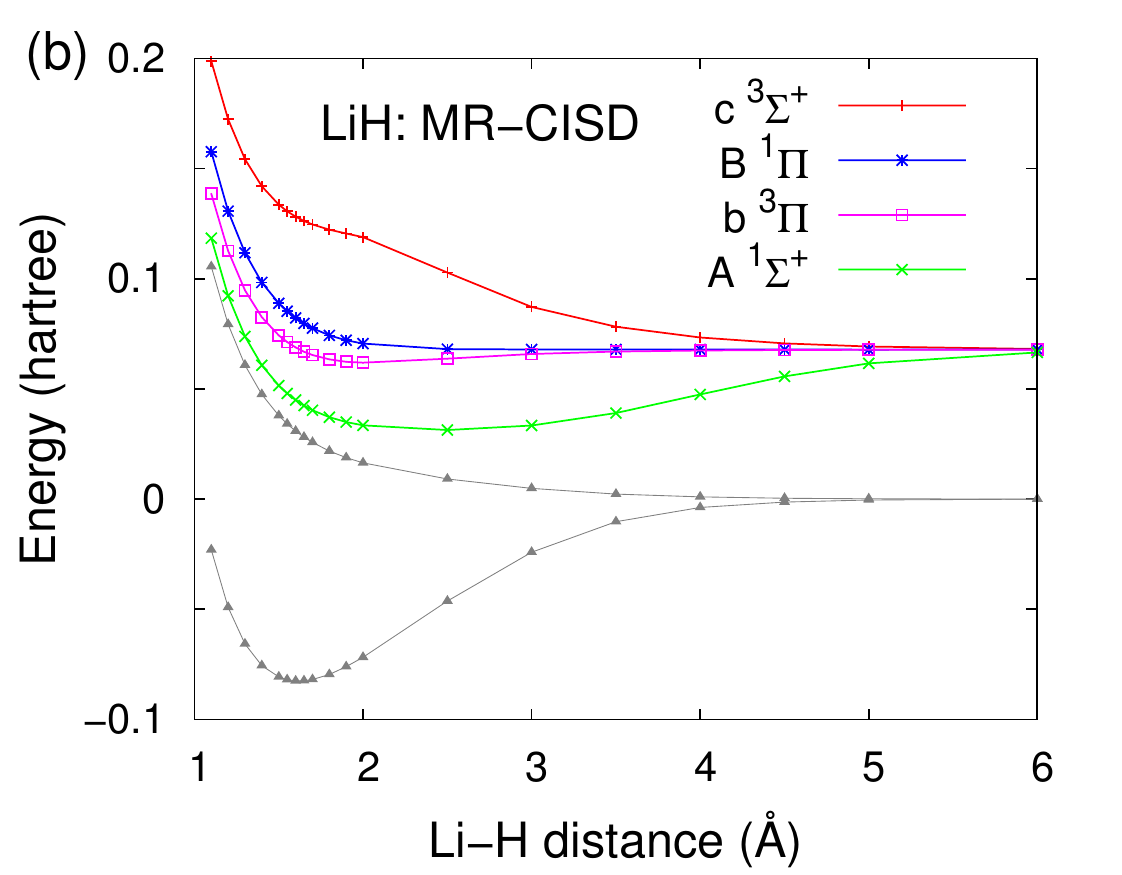}
    \caption{
Potential energy curves of $A^{1}\Sigma^{+}$, $B^{1}\Pi$,
$b^{3}\Pi$, $c^{1}\Sigma^{+}$ excited states of LiH.
(a) the present electron wave packet calculation.
(b) multi-reference configuration-interaction with single and double excitations
with the cc-pVDZ basis set.
The shifts of absolute energies 
are identical to those indicated in the caption of Fig.1.
} 
\end{figure}

Figure 3(a) presents the resultant potential energy curves 
for the excited $A{}^{1}\Sigma^{+}$,
$B{}^{1}\Pi$, $b{}^{3}\Pi$, and $c{}^{3}\Sigma^{+}$ states of LiH. 
Despite the simplicity of calculation, the potential curves in
Fig. 3(a) exhibit semi-quantitative accuracy 
in comparison with the MR-CISD reference displayed
in Fig. 3(b). 
For the $A^{1}\Sigma^{+}$ state, we have double-checked the results
by constructing a two-dimensional potential energy surface 
for the motions of two electron WPs corresponding to the H 1s
and the Li 2s, 
because their overlap is apparent in Fig. 2(a)
for the $X^{1}\Sigma^{+}$ state.
However,
this additional calculation
did not affect notably the results
in Fig. 3(a). 
We have confirmed the reason that 
these degrees of freedom are well separated in terms of both the shape and curvature 
of the potential surface,
as the H 1s WP is much more tightly
bound around the nucleus than the Li 2s WP.

\begin{figure}[t]
    \centering
        \includegraphics[width=0.45\textwidth]{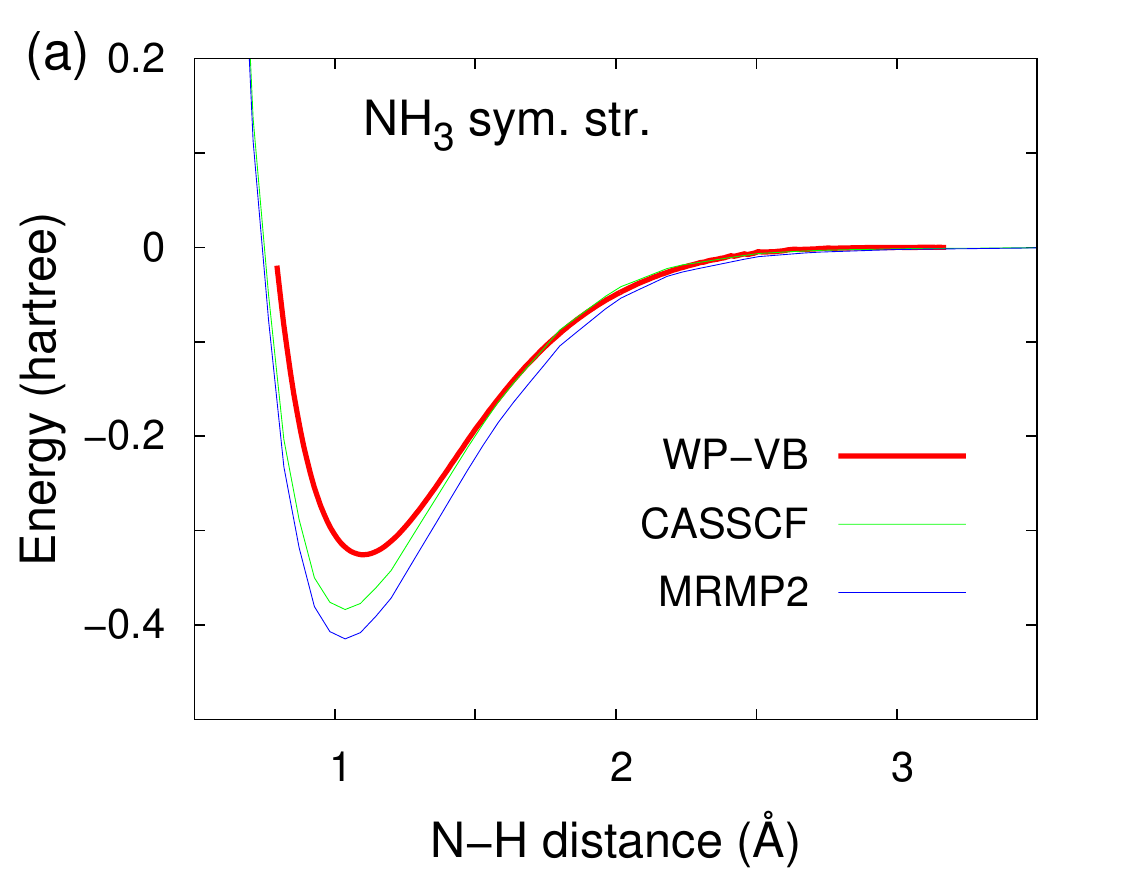}
        \includegraphics[width=0.35\textwidth]{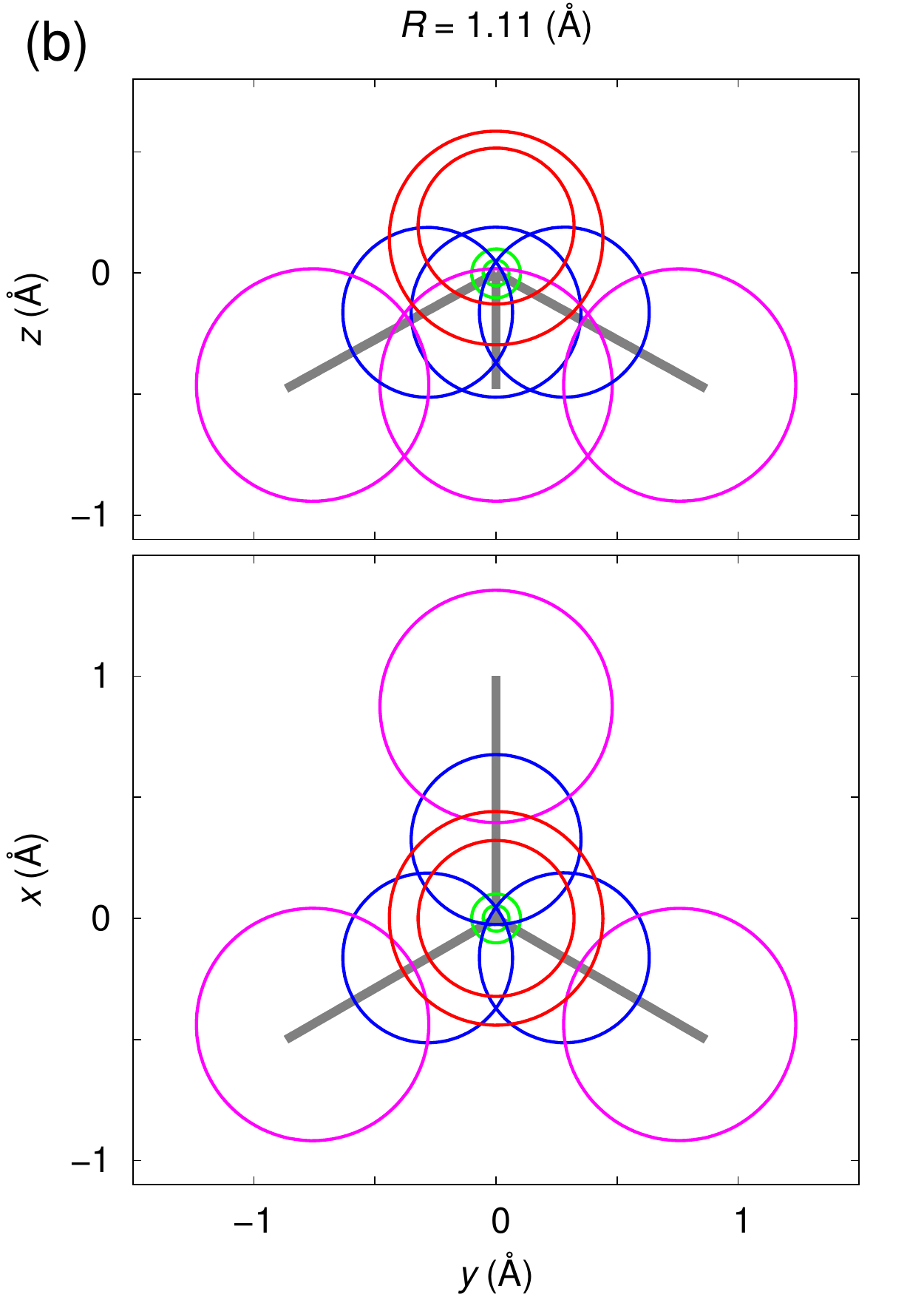}
    \caption{
    Potential energy curves and electron wave packets for NH$_3$.
    (a) potential energy curves along the 
    N-H symmetric stretch coordinate of NH$_3$, 
    comparing the present model of electron wave packets with valence-bond coupling (WP-VB),
    complete active space self-consistent field (CASSCF), and
    multi-reference second-order M{\o}ller-Plesset perturbation theory (MPMP2),
    the latter two with the cc-pVDZ basis set.
    (b) electron wave packets for NH$_3$ with the radius of
    circles representing the wave packet widths ($\rho$).
    }
\end{figure}

It might appear that the success of the spherical WP-VB 
was just because only 1s and 2s AOs are involved in LiH.
However, we demonstrate in Fig. 4 that the potential energy curve
for a polyatomic molecule, NH$_3$, is accurately described.
The conventional description of this molecule involves nitrogen 2p AOs
for the NH bonds and lone-pair electrons.
Figure 4(b) shows that these are properly described by the spherical WPs.
The key is, as noted for Fig. 2, that the floating WPs
account for the polarization of electronic wave function,
for which
the conventional MO methods require AOs of higher angular momenta
as the AOs are clamped at the nuclear centers.
Their nodal structures are not essential 
for generally nodeless \textit{total} electronic wave functions in
the ground state.
As demonstrated in Fig. 3, the excited states can be studied
with the electron WP dynamics.

Now the major bottleneck for 
applications to polyatomic molecules is not fundamental but
mostly technical.
In order to improve quantitative accuracy,
the first task will be to employ more flexible WPs than the spherical ones.
This will complicate the evaluation of two-electron integrals, but
we have already applied ellipsoidal WPs
to nuclear WP simulation of liquid water \cite{Ono12}.
Another task would be extension to multi-configuration spin-couplings
that will be needed for some chemical reactions \cite{Shaik08}. 
Nonetheless,
only a few spin-coupling patterns will suffice in practice for essential
picture, especially due to the flexibility of floating and breathing WPs.
Even in such cases, with use of single electron WPs for
the spatial part, 
the corpuscular picture of electrons will remain viable.

\section*{Acknowledgment}
This work was supported by KAKENHI No. 26248009 and 26620007.


\begin{thebibliography}{30}
\expandafter\ifx\csname natexlab\endcsname\relax\def\natexlab#1{#1}\fi
\expandafter\ifx\csname bibnamefont\endcsname\relax
  \def\bibnamefont#1{#1}\fi
\expandafter\ifx\csname bibfnamefont\endcsname\relax
  \def\bibfnamefont#1{#1}\fi
\expandafter\ifx\csname citenamefont\endcsname\relax
  \def\citenamefont#1{#1}\fi
\expandafter\ifx\csname url\endcsname\relax
  \def\url#1{\texttt{#1}}\fi
\expandafter\ifx\csname urlprefix\endcsname\relax\def\urlprefix{URL }\fi
\providecommand{\bibinfo}[2]{#2}
\providecommand{\eprint}[2][]{\url{#2}}

\bibitem[{\citenamefont{Szabo and Ostlund}(1996)}]{SzaboOstlund}
\bibinfo{author}{\bibfnamefont{A.}~\bibnamefont{Szabo}} \bibnamefont{and}
  \bibinfo{author}{\bibfnamefont{N.~S.} \bibnamefont{Ostlund}},
  \emph{\bibinfo{title}{Modern Quantum Chemistry}} (\bibinfo{publisher}{Dover},
  \bibinfo{address}{New York}, \bibinfo{year}{1996}).

\bibitem[{\citenamefont{McWeeny}(1992)}]{McWeeny92}
\bibinfo{author}{\bibfnamefont{R.}~\bibnamefont{McWeeny}},
  \emph{\bibinfo{title}{Methods of Molecular Quantum Mechanics}}
  (\bibinfo{publisher}{Academic}, \bibinfo{address}{London},
  \bibinfo{year}{1992}).

\bibitem[{\citenamefont{Parr and Yang}(1989)}]{Parr89}
\bibinfo{author}{\bibfnamefont{R.~G.} \bibnamefont{Parr}} \bibnamefont{and}
  \bibinfo{author}{\bibfnamefont{W.}~\bibnamefont{Yang}},
  \emph{\bibinfo{title}{Density Functional Theory of Atoms and Molecules}}
  (\bibinfo{publisher}{Oxford University}, \bibinfo{address}{Oxford},
  \bibinfo{year}{1989}).

\bibitem[{\citenamefont{Fukui}(1971)}]{Fukui71}
\bibinfo{author}{\bibfnamefont{K.}~\bibnamefont{Fukui}}, \bibinfo{journal}{Acc.
  Chem. Res.} \textbf{\bibinfo{volume}{4}}, \bibinfo{pages}{57}
  (\bibinfo{year}{1971}).

\bibitem[{\citenamefont{Woodward and Hoffmann}(1969)}]{Woodward69}
\bibinfo{author}{\bibfnamefont{R.~B.} \bibnamefont{Woodward}} \bibnamefont{and}
  \bibinfo{author}{\bibfnamefont{R.}~\bibnamefont{Hoffmann}},
  \bibinfo{journal}{Angew. Chem. Int. Ed. Engl.} \textbf{\bibinfo{volume}{11}},
  \bibinfo{pages}{781} (\bibinfo{year}{1969}).

\bibitem[{\citenamefont{Itatani et~al.}(2004)\citenamefont{Itatani, Levesque,
  Zeldler, Niikura, P{\'e}pin, Kleffer, Corkum, and Villeneuve}}]{Itatani04}
\bibinfo{author}{\bibfnamefont{J.}~\bibnamefont{Itatani}},
  \bibinfo{author}{\bibfnamefont{J.}~\bibnamefont{Levesque}},
  \bibinfo{author}{\bibfnamefont{D.}~\bibnamefont{Zeldler}},
  \bibinfo{author}{\bibfnamefont{H.}~\bibnamefont{Niikura}},
  \bibinfo{author}{\bibfnamefont{H.}~\bibnamefont{P{\'e}pin}},
  \bibinfo{author}{\bibfnamefont{J.~C.} \bibnamefont{Kleffer}},
  \bibinfo{author}{\bibfnamefont{P.~B.} \bibnamefont{Corkum}},
  \bibnamefont{and} \bibinfo{author}{\bibfnamefont{D.~M.}
  \bibnamefont{Villeneuve}}, \bibinfo{journal}{Nature}
  \textbf{\bibinfo{volume}{432}}, \bibinfo{pages}{867} (\bibinfo{year}{2004}).

\bibitem[{\citenamefont{Smirnova et~al.}(2009)\citenamefont{Smirnova, Mairesse,
  Patchkovskii, Dudovich, Villeneuve, Corkum, and Ivanov}}]{Smirnova09}
\bibinfo{author}{\bibfnamefont{O.}~\bibnamefont{Smirnova}},
  \bibinfo{author}{\bibfnamefont{Y.}~\bibnamefont{Mairesse}},
  \bibinfo{author}{\bibfnamefont{S.}~\bibnamefont{Patchkovskii}},
  \bibinfo{author}{\bibfnamefont{N.}~\bibnamefont{Dudovich}},
  \bibinfo{author}{\bibfnamefont{D.}~\bibnamefont{Villeneuve}},
  \bibinfo{author}{\bibfnamefont{P.}~\bibnamefont{Corkum}}, \bibnamefont{and}
  \bibinfo{author}{\bibfnamefont{M.~Y.} \bibnamefont{Ivanov}},
  \bibinfo{journal}{Nature} \textbf{\bibinfo{volume}{460}},
  \bibinfo{pages}{972} (\bibinfo{year}{2009}).

\bibitem[{\citenamefont{Yamazaki et~al.}(2015)\citenamefont{Yamazaki, Oishi,
  Nakazawa, Zhu, and Takahashi}}]{Yamazaki15}
\bibinfo{author}{\bibfnamefont{M.}~\bibnamefont{Yamazaki}},
  \bibinfo{author}{\bibfnamefont{K.}~\bibnamefont{Oishi}},
  \bibinfo{author}{\bibfnamefont{H.}~\bibnamefont{Nakazawa}},
  \bibinfo{author}{\bibfnamefont{C.}~\bibnamefont{Zhu}}, \bibnamefont{and}
  \bibinfo{author}{\bibfnamefont{M.}~\bibnamefont{Takahashi}},
  \bibinfo{journal}{Phys. Rev. Lett.} \textbf{\bibinfo{volume}{114}},
  \bibinfo{pages}{103005} (\bibinfo{year}{2015}).

\bibitem[{\citenamefont{Ohno}(2004)}]{Ohno04}
\bibinfo{author}{\bibfnamefont{K.}~\bibnamefont{Ohno}}, \bibinfo{journal}{Bull.
  Chem. Soc. Jpn.} \textbf{\bibinfo{volume}{77}}, \bibinfo{pages}{887}
  (\bibinfo{year}{2004}).

\bibitem[{\citenamefont{Shaik and Hiberty}(2008)}]{Shaik08}
\bibinfo{author}{\bibfnamefont{S.~S.} \bibnamefont{Shaik}} \bibnamefont{and}
  \bibinfo{author}{\bibfnamefont{P.~C.} \bibnamefont{Hiberty}},
  \emph{\bibinfo{title}{A Chemist's Guide to Valence Bond Theory}}
  (\bibinfo{publisher}{Wiley}, \bibinfo{address}{New Jersey},
  \bibinfo{year}{2008}).

\bibitem[{\citenamefont{Pauling}(1960)}]{Pauling60}
\bibinfo{author}{\bibfnamefont{L.}~\bibnamefont{Pauling}},
  \emph{\bibinfo{title}{The Nature of the Chemical Bond}}
  (\bibinfo{publisher}{Cornell University Press}, \bibinfo{address}{New York},
  \bibinfo{year}{1960}).

\bibitem[{\citenamefont{Ando}(2009)}]{Ando09}
\bibinfo{author}{\bibfnamefont{K.}~\bibnamefont{Ando}}, \bibinfo{journal}{Bull.
  Chem. Soc. Jpn.} \textbf{\bibinfo{volume}{82}}, \bibinfo{pages}{975}
  (\bibinfo{year}{2009}).

\bibitem[{\citenamefont{Ando}(2012)}]{Ando12}
\bibinfo{author}{\bibfnamefont{K.}~\bibnamefont{Ando}}, \bibinfo{journal}{Chem.
  Phys. Lett.} \textbf{\bibinfo{volume}{523}}, \bibinfo{pages}{134}
  (\bibinfo{year}{2012}).

\bibitem[{\citenamefont{Kling and Vrakking}(2008)}]{Kling08}
\bibinfo{author}{\bibfnamefont{M.~F.} \bibnamefont{Kling}} \bibnamefont{and}
  \bibinfo{author}{\bibfnamefont{M.~J.~J.} \bibnamefont{Vrakking}},
  \bibinfo{journal}{Ann. Rev. Phys. Chem.} \textbf{\bibinfo{volume}{59}},
  \bibinfo{pages}{463} (\bibinfo{year}{2008}).

\bibitem[{\citenamefont{Krausz and Ivanov}(2009)}]{Krausz09}
\bibinfo{author}{\bibfnamefont{F.}~\bibnamefont{Krausz}} \bibnamefont{and}
  \bibinfo{author}{\bibfnamefont{M.}~\bibnamefont{Ivanov}},
  \bibinfo{journal}{Rev. Mod. Phys.} \textbf{\bibinfo{volume}{81}},
  \bibinfo{pages}{163} (\bibinfo{year}{2009}).

\bibitem[{\citenamefont{Kato et~al.}(2012)\citenamefont{Kato, Oyamada, Kono,
  and Koseki}}]{Kato12}
\bibinfo{author}{\bibfnamefont{T.}~\bibnamefont{Kato}},
  \bibinfo{author}{\bibfnamefont{T.}~\bibnamefont{Oyamada}},
  \bibinfo{author}{\bibfnamefont{H.}~\bibnamefont{Kono}}, \bibnamefont{and}
  \bibinfo{author}{\bibfnamefont{S.}~\bibnamefont{Koseki}},
  \bibinfo{journal}{Prog. Theor. Phys. Suppl.} \textbf{\bibinfo{volume}{196}},
  \bibinfo{pages}{16} (\bibinfo{year}{2012}).

\bibitem[{\citenamefont{Vacher et~al.}(2014)\citenamefont{Vacher,
  Mendive-Tapia, Bearpark, and Robb}}]{Vacher14}
\bibinfo{author}{\bibfnamefont{M.}~\bibnamefont{Vacher}},
  \bibinfo{author}{\bibfnamefont{D.}~\bibnamefont{Mendive-Tapia}},
  \bibinfo{author}{\bibfnamefont{M.~J.} \bibnamefont{Bearpark}},
  \bibnamefont{and} \bibinfo{author}{\bibfnamefont{M.~A.} \bibnamefont{Robb}},
  \bibinfo{journal}{Theor. Chem. Acc.} \textbf{\bibinfo{volume}{133}},
  \bibinfo{pages}{1505} (\bibinfo{year}{2014}).

\bibitem[{\citenamefont{Oliveira et~al.}(2015)\citenamefont{Oliveira, Mignolet,
  Kus, Papadopoulos, Remacle, and Verstraete}}]{Oliveira15}
\bibinfo{author}{\bibfnamefont{M.~J.~T.} \bibnamefont{Oliveira}},
  \bibinfo{author}{\bibfnamefont{B.}~\bibnamefont{Mignolet}},
  \bibinfo{author}{\bibfnamefont{T.}~\bibnamefont{Kus}},
  \bibinfo{author}{\bibfnamefont{T.~A.} \bibnamefont{Papadopoulos}},
  \bibinfo{author}{\bibfnamefont{F.}~\bibnamefont{Remacle}}, \bibnamefont{and}
  \bibinfo{author}{\bibfnamefont{M.~J.} \bibnamefont{Verstraete}},
  \bibinfo{journal}{J. Chem. Theory Comput.} \textbf{\bibinfo{volume}{11}},
  \bibinfo{pages}{2221} (\bibinfo{year}{2015}).

\bibitem[{\citenamefont{Frost}(1967)}]{Frost67}
\bibinfo{author}{\bibfnamefont{A.~A.} \bibnamefont{Frost}},
  \bibinfo{journal}{J. Chem. Phys.} \textbf{\bibinfo{volume}{47}},
  \bibinfo{pages}{3707} (\bibinfo{year}{1967}).

\bibitem[{\citenamefont{Nakatsuji et~al.}(1978)\citenamefont{Nakatsuji,
  Kanayama, Harada, and Yonezawa}}]{Nakatsuji78}
\bibinfo{author}{\bibfnamefont{H.}~\bibnamefont{Nakatsuji}},
  \bibinfo{author}{\bibfnamefont{S.}~\bibnamefont{Kanayama}},
  \bibinfo{author}{\bibfnamefont{S.}~\bibnamefont{Harada}}, \bibnamefont{and}
  \bibinfo{author}{\bibfnamefont{T.}~\bibnamefont{Yonezawa}},
  \bibinfo{journal}{J. Am. Chem. Soc.} \textbf{\bibinfo{volume}{100}},
  \bibinfo{pages}{7528} (\bibinfo{year}{1978}).

\bibitem[{\citenamefont{Shull and Ebbing}(1958)}]{Shull58}
\bibinfo{author}{\bibfnamefont{H.}~\bibnamefont{Shull}} \bibnamefont{and}
  \bibinfo{author}{\bibfnamefont{D.~D.} \bibnamefont{Ebbing}},
  \bibinfo{journal}{J. Chem. Phys.} \textbf{\bibinfo{volume}{28}},
  \bibinfo{pages}{866} (\bibinfo{year}{1958}).

\bibitem[{\citenamefont{Reeves}(1963)}]{Reeves63}
\bibinfo{author}{\bibfnamefont{C.~M.} \bibnamefont{Reeves}},
  \bibinfo{journal}{J. Chem. Phys.} \textbf{\bibinfo{volume}{39}},
  \bibinfo{pages}{1} (\bibinfo{year}{1963}).

\bibitem[{\citenamefont{Arickx et~al.}(1986)\citenamefont{Arickx, Broeckhove,
  Kesteloot, Lathouwers, and van Leuven}}]{Arickx86}
\bibinfo{author}{\bibfnamefont{F.}~\bibnamefont{Arickx}},
  \bibinfo{author}{\bibfnamefont{J.}~\bibnamefont{Broeckhove}},
  \bibinfo{author}{\bibfnamefont{E.}~\bibnamefont{Kesteloot}},
  \bibinfo{author}{\bibfnamefont{L.}~\bibnamefont{Lathouwers}},
  \bibnamefont{and} \bibinfo{author}{\bibfnamefont{P.}~\bibnamefont{van
  Leuven}}, \bibinfo{journal}{Chem. Phys. Lett.}
  \textbf{\bibinfo{volume}{128}}, \bibinfo{pages}{310} (\bibinfo{year}{1986}).

\bibitem[{\citenamefont{Kuratsuji}(1981)}]{Kuratsuji81}
\bibinfo{author}{\bibfnamefont{H.}~\bibnamefont{Kuratsuji}},
  \bibinfo{journal}{Prog. Theor. Phys.} \textbf{\bibinfo{volume}{65}},
  \bibinfo{pages}{224} (\bibinfo{year}{1981}).

\bibitem[{\citenamefont{Ando}(2014)}]{Ando14}
\bibinfo{author}{\bibfnamefont{K.}~\bibnamefont{Ando}}, \bibinfo{journal}{Chem.
  Phys. Lett.} \textbf{\bibinfo{volume}{591}}, \bibinfo{pages}{179}
  (\bibinfo{year}{2014}).

\bibitem[{\citenamefont{Tsue}(1992)}]{Tsue92}
\bibinfo{author}{\bibfnamefont{Y.}~\bibnamefont{Tsue}}, \bibinfo{journal}{Prog.
  Theor. Phys.} \textbf{\bibinfo{volume}{88}}, \bibinfo{pages}{911}
  (\bibinfo{year}{1992}).

\bibitem[{\citenamefont{Schmidt et~al.}(1993)\citenamefont{Schmidt, Baldridge,
  Boatz, Elbert, Gordon, Jensen, Koseki, Matsunaga, Nguyen, Su
  et~al.}}]{GAMESS}
\bibinfo{author}{\bibfnamefont{M.~W.} \bibnamefont{Schmidt}},
  \bibinfo{author}{\bibfnamefont{K.~K.} \bibnamefont{Baldridge}},
  \bibinfo{author}{\bibfnamefont{J.~A.} \bibnamefont{Boatz}},
  \bibinfo{author}{\bibfnamefont{S.~T.} \bibnamefont{Elbert}},
  \bibinfo{author}{\bibfnamefont{M.~S.} \bibnamefont{Gordon}},
  \bibinfo{author}{\bibfnamefont{J.~H.} \bibnamefont{Jensen}},
  \bibinfo{author}{\bibfnamefont{S.}~\bibnamefont{Koseki}},
  \bibinfo{author}{\bibfnamefont{N.}~\bibnamefont{Matsunaga}},
  \bibinfo{author}{\bibfnamefont{K.~A.} \bibnamefont{Nguyen}},
  \bibinfo{author}{\bibfnamefont{S.~J.} \bibnamefont{Su}},
  \bibnamefont{et~al.}, \bibinfo{journal}{J. Comp. Chem.}
  \textbf{\bibinfo{volume}{14}}, \bibinfo{pages}{1347} (\bibinfo{year}{1993}).

\bibitem[{\citenamefont{Klauder and Skagerstam}(1985)}]{Klauder84}
\bibinfo{editor}{\bibfnamefont{J.~R.} \bibnamefont{Klauder}} \bibnamefont{and}
  \bibinfo{editor}{\bibfnamefont{B.~S.} \bibnamefont{Skagerstam}}, eds.,
  \emph{\bibinfo{title}{Coherent States: Applications in Physics and
  Mathematical Physics}} (\bibinfo{publisher}{World Scientific},
  \bibinfo{address}{Singapore}, \bibinfo{year}{1985}).

\bibitem[{\citenamefont{Kay}(2005)}]{Kay05}
\bibinfo{author}{\bibfnamefont{K.~G.} \bibnamefont{Kay}},
  \bibinfo{journal}{Ann. Rev. Phys. Chem.} \textbf{\bibinfo{volume}{56}},
  \bibinfo{pages}{255} (\bibinfo{year}{2005}).

\bibitem[{\citenamefont{Ono and Ando}(2012)}]{Ono12}
\bibinfo{author}{\bibfnamefont{J.}~\bibnamefont{Ono}} \bibnamefont{and}
  \bibinfo{author}{\bibfnamefont{K.}~\bibnamefont{Ando}}, \bibinfo{journal}{J.
  Chem. Phys.} \textbf{\bibinfo{volume}{137}}, \bibinfo{pages}{174503}
  (\bibinfo{year}{2012}).

\end{thebibliography}
\end{document}